\documentclass[aps,prb,showpacs,twocolumn,floats,superscriptaddress,10pt]{revtex4-1}
\usepackage{graphicx} \usepackage{dcolumn}
\usepackage{bm}
\usepackage{color}
\usepackage{amssymb}
\usepackage{amsmath}
\usepackage{simplewick}
\usepackage{array}
\usepackage{appendix}

\RequirePackage[
   hyperindex,colorlinks,bookmarksnumbered,
   plainpages=true,pdfstartview=FitH]{hyperref}
\hypersetup{linkcolor=blue,urlcolor=blue,citecolor=blue}
\usepackage{hyperref}

\usepackage{ulem}
\renewcommand{\emph}[1]{\textit{#1}}

\newcommand{\Eq}[1]{Eq.\,(\ref{#1})}
\newcommand{\Eqs}[1]{Eqs.\,(\ref{#1})}

\newcommand{\vF}{v_{\rm F}}
\newcommand{\eF}{\varepsilon_{\rm F}}
\newcommand{\Bimp}{B}
\newcommand{\Tk}{T_{\rm K}}
\newcommand{\half}{{\scriptscriptstyle {1 \! / \! 2}}}
\newcommand{\TkFS}{T^{\rm FS}_{\rm K}}
\newcommand{\Tkloc}{T^{\rm loc}_{\rm K}}
\newcommand{\Sztot}{S_z^{\rm tot}}

\newcommand{\discrete}{{\rm disc}}
\newcommand{\Tksc}{T^{\rm sc}_{\rm K}}
\newcommand{\Ezsigma}{\varepsilon^{\rm Z}_\sigma}
\newcommand{\imp}{{\rm imp}}
\newcommand{\magnetic}{{\rm mag}}
\newcommand{\bsigma}{\bar \sigma}
\newcommand{\chifree}{\chi^{\rm free}}
\newcommand{\chiimp}{\chi^{\rm imp}}

\newcommand{\chiFS}{\chi^{\rm FS}}
\newcommand{\chiloc}{\chi^{\rm loc}}
\newcommand{\chisc}{\chi^{\rm sc}}
\newcommand{\Cimp}{C^{\rm imp}}
\newcommand{\barn}{\delta \bar n}
\newcommand{\tdelta}{\tilde \delta}

\newcommand{\T}{\mathcal{T}}

\newcommand{\tTin}{\tilde {\mathcal{T}}^{\rm in}}
\newcommand{\tTel}{\tilde {\mathcal{T}}^{\rm el}}
\newcommand{\tSigma}{\tilde \Sigma^{R}}
\newcommand{\tSigmaone}{\tilde \Sigma^{R,1}}
\newcommand{\tSigmatwo}{\tilde \Sigma^{R,2}}
\newcommand{\tSigmathree}{\tilde \Sigma^{R,3}}
\newcommand{\Jk}{J_{\rm K}}

\newcommand{\Hin}{H_{\rm int}}
\newcommand{\HtwoMF}{H_2^{\rm MF}}
\newcommand{\HthreeMF}{H_3^{\rm MF}}
\newcommand{\Hone}{H_1}
\newcommand{\Htwo}{H_2}
\newcommand{\Hthree}{H_3}
\newcommand{\Hloc}{H_{\rm loc}}
\newcommand{\Hfp}{H_{\rm fp}}
\newcommand{\qph}{\quad \phantom{.}}
\newcommand{\qqph}{\qquad \phantom{.}}

\newcommand{\FLT}{{\rm FLT}}

\begin{document}
\title{Equilibrium Fermi-liquid coefficients for the fully 
screened $N$-channel Kondo model}
\author{M. Hanl}
\author{A. Weichselbaum}
\author{J. von Delft}
\affiliation{Physics Department, Arnold Sommerfeld Center for Theoretical Physics and Center for NanoScience, Ludwig-Maximilians-Universit\"at M\"unchen, 80333 M\"unchen, Germany}
\author{M. Kiselev}
\affiliation{The Abdus Salam International Centre for Theoretical Physics,
Strada Costiera 11, I-34151 Trieste, Italy}
\date{May 27, 2014}

\begin{abstract}
  We analytically and numerically compute three equilibrium
  Fermi-liquid coefficients of the fully screened $N$-channel Kondo
  model, namely $c_B$, $c_T$ and $c_\varepsilon$, characterizing the
  magnetic field and temperature-dependence of the resisitivity, and
  the curvature of the equilibrium Kondo resonance, respectively.  We
  present a compact, unified derivation of the $N$-dependence of these
  coefficients, combining elements from various
  previous treatments of this model.  We numerically compute
  these coefficients using the numerical renormalization group, with
  non-Abelian symmetries implemented explicitly, finding agreement
  with Fermi-liquid predictions on the order of 5\% or better.
\end{abstract}

\maketitle

\section{Introduction} 

The Kondo effect was first observed, in the 1930s, for iron
impurities in gold and silver \cite{deHaas1934,deHaas1936}, as an
anomalous rise in the resistivity with decreasing
temperature. Kondo\cite{Kondo1964} showed that this effect is caused
by an antiferromagnetic exchange coupling between the localized
magnetic impurity spins and the spins of the delocalized conduction
electrons \cite{Kondo1964}, and based his arguments on a
spin-$\frac{1}{2}$, one-band model.  While this model undoubtedly
captures the essential physics correctly in a qualitative way, it
has recently been shown\cite{Costi2009,Hanl2013} that a
quantitatively correct description of the Kondo physics of dilute Fe
impurities in Au or Ag requires a fully screened Kondo model
involving three channels and a spin-$\frac{3}{2}$ impurity.  This
conclusion was based on a comparison of temperature and magnetic
field dependent transport
measurements\cite{Mallet2006,Costi2009,Hanl2013} to theoretical
predictions for fully screened Kondo models with channel number $N$
and local spin $S$ related by $N = 2S$, with  $N = 3$
yielding much better agreement than $N=1$ or 2. 

The theoretical results in Ref.~\onlinecite{Hanl2013} were obtained
using the numerical renormalization group
(NRG),\cite{Wilson1975,Krishnamurthy1980,Weichselbaum2007,Bulla2008}
and for $N=3$ various non-Abelian
symmetries\cite{Weichselbaum2012b,Hanl2013}, such as
SU(2)$\times$U(1)$\times$SU$(N)$, had to be exploited to
achieve reliable results at finite magnetic field. The technology
for implementing non-Abelian symmetries with $N > 2$ in NRG
calculations has been developed only
recently.\cite{Weichselbaum2012b,Moca2012} Given the complexity of
such calculations, it is desirable to benchmark their quality by
comparing their predictions to exact results. The motivation for the
present paper was to perform such a comparison for the low-energy
Fermi-liquid behavior of fully screened Kondo models, as elaborated upon
below.

All fully screened Kondo models feature a ground state in which the
impurity spin is screened by the conduction electrons into a spin
singlet. The low-energy behavior of these models can be described by
a phenomenological Fermi-liquid theory (FLT) formulated in terms of
the phase shift  experienced by
conduction electrons that scatter elastically off the screened
singlet.  Such a description was first devised for the simplest case
of $N=1$ by Nozi\`eres\cite{Nozieres1974,Nozieres1974a} in 1974,
and generalized to the case of arbitrary $N$ by Nozi\`eres and
Blandin (NB)\cite{Nozieres1980} in 1980. Their results were
confirmed and elaborated by various authors and methods, including
NRG,\cite{Wilson1975,Krishnamurthy1980,Cragg1978,Cragg1979,Cragg1980,Pang1992,N12}
field-theoretic calculations,\cite{Yoshimori,Zaw} the Bethe
Ansatz,\cite{WiT,Andrei} conformal field theory
(CFT),\cite{A90,AL93} renormalized perturbation theory,\cite{H10}
and reformulations\cite{Pustilnik2001,PG,Glazman2005}
and generalizations\cite{Mora09a,Mora09b,Mitchell2012} of Nozi\`eres' approach 
in the context of Kondo quantum dots.

In the present paper, we focus on three particular Fermi-liquid
coefficients, $c_B$, $c_T$ and $c_\varepsilon$,
characterizing the leading dependence of the resistivity on magnetic
field ($B$) and temperature ($T$), and the curvature of the
equilibrium Kondo resonance as function of excitation energy
$(\varepsilon$), respectively.  Explicit formulas for all three of
these coefficients are available in the literature for $N=1$, but
for general $N$ only for the case of $c_T$. Given the wealth of
previous studies of fully-screened Kondo models, the lack of
corresponding formulas for $c_B$ and $c_\varepsilon$ was somewhat
unexpected.  Thus, we offer here a unified derivation of all three
Fermi-liquid coefficients, $c_T$, $c_B$ and $c_\varepsilon$. We
follow the strategy which Affleck and Ludwig (AL)\cite{AL93} have
used to reproduce Nozi\`eres' results\cite{Nozieres1974} for $N=1$,
namely doing perturbation theory in the leading irrelevant operator,
and generalize it to the case of arbitrary $N$. Our formulation of
this strategy follows that used by Pustilnik and Glazman
(PG)\cite{PG} for their discussion of Kondo quantum dots.  While all
pertinent ideas used here can be found in the literature, we hope
that our rather compact way of combining them will be found useful.

For our numerical work, we faced two challenges: First, the
complexity of the numerical calculations increases rapidly with
increasing $N$; this was dealt with by exploiting non-Abelian
symmetries.  Second, numerical calculations do not achieve the scaling
limit that is implicitely presumed in analytical calculations; its
absence was compensated by using suitable definitions of the Kondo
temperature, following Ref.~\onlinecite{Hanl2013a}.

The paper is organized as follows. In Sec.~\ref{sec:model} we define
the model and summarize our key results for the Fermi-liquid
coefficients $c_B$, $c_T$ and
$c_\varepsilon$. Section~\ref{sec:Fermi-liquid} compactly summarizes
relevant elements of FLT and uses them to calculate
these coefficients.  Section~\ref{sec:NRG} describes our numerical work
and results. Section~\ref{sec:conclusions}
summarizes our conclusions.

\section{Model and Main Results}
\label{sec:model}

The fully-screened Kondo model for $N$ conduction
bands coupled to a single magnetic impurity at the origin
is defined by the Hamiltonian $H=H_0 + \Hloc$, with 
\begin{subequations}
\label{eq:Kondomodel}
\begin{eqnarray}
\label{eq:define-H0}
H_0 & = & 
\sum_{km\sigma}\xi_k c^{\dagger}_{km\sigma}c_{km\sigma}  \; , 
\\
\Hloc & = & 
\label{eq:define-Hloc}
J_{\rm K} \sum_{kk'm\sigma\sigma'}
c_{km\sigma}^{\dag}\frac{\vec \tau_{\sigma\sigma'}}{2} 
c_{k'm\sigma'}\vec S  - BS_z .
\end{eqnarray}
\end{subequations}
Here $H_0$ describes $N$
channels of free conduction electrons, with
spin index $\sigma=(+,-)=(\uparrow,\downarrow)$ and channel index $m =
1, \dots, N$. We take the dispersion $\xi_k = \varepsilon_k - \eF$ to
be linear and symmetric around the Fermi energy, $\xi_k = k \hbar
\vF$.  Each channel has exchange coupling $\Jk$ to a local SU(2) spin
of size $S= N/2$ with spin operators $\vec S$, and $B$ describes a
local Zeeman field in the $z$-direction (we use units $g\mu_B=1$).
The overall symmetry of the model\cite{Pang1992} is SU(2)$\times$
Sp$(2N)$ for $B=0$, and U(1)$\times$Sp$(2N)$ for $B\neq 0$ (see
Sec.~\ref{sec:NRG-details} for details). The model is characterized by
a low-energy scale, the Kondo temperature, $\Tk \sim \tilde D
\exp\left[-1/(\nu \Jk)\right]$, where $\nu$ is the density of states
per channel and spin species and $\tilde D$ is of the order of the
conduction electron bandwidth.

For a disordered metal containing a dilute concentration of magnetic impurities,
the magnetic-impurity  contribution to the resisitivity has the form\cite{Hanl2013,Drude}
\begin{equation}
\rho (T,B)\propto \int d\varepsilon 
\bigl(- \partial_\varepsilon  f(\varepsilon,T)\bigr)
\sum_{m\sigma}A_{m\sigma}(\varepsilon,T,B) \; . 
\label{eq:G(T,B)}
\end{equation}
Here $f(\varepsilon,T)$ is the Fermi function, and the impurity
spectral function $A_{m\sigma}(\varepsilon) = -\frac{1}{\pi}{\rm Im}
\T_{m\sigma}(\varepsilon) $ is the imaginary part of the $T$ matrix
$\T_{m\sigma}(\varepsilon)$ describing scattering off a
magnetic impurity. The latter is defined through\cite{Costi2000,Rosch2003}
\begin{align}
{\cal G}_{m\sigma,{\mathbf k},{\mathbf k'}}^c(\varepsilon)&={\cal G}_{m\sigma,\mathbf k}^0(\varepsilon)\delta ({\mathbf k}-{\mathbf k'})\nonumber \\
&+{\cal G}_{m\sigma,\mathbf k}^0(\varepsilon){\cal T}_{m\sigma}(\varepsilon) {\cal G}_{m\sigma,\mathbf k'}^0(\varepsilon) \; ,
\label{eq:define-T-matrix} 
\end{align}
with ${\cal G}_{m\sigma,{\mathbf k},{\mathbf k'}}^c$ and ${\cal
 G}_{m\sigma,\mathbf k}^0$ the full and bare conduction electron
Green's functions, respectively.  [For a Kondo quantum dot tuned such
that the low-energy physics is described by \Eq{eq:Kondomodel}, the
conductance $G$ through the dot has a form similar to \Eq{eq:G(T,B)},
with $\rho$ replaced by $G$.\cite{PG}]

As mentioned in the Introduction, the ground state of the
fully screened Kondo model is a spin singlet, and the regime
of low-energy excitations below $\Tk$ shows Fermi-liquid
behavior.\cite{Nozieres1974,Nozieres1980} One characteristic
Fermi-liquid property is that 
the leading dependence of the $T$ matrix on
its arguments, when they are small relative to $\Tk$,
is quadratic,
\begin{eqnarray}
\label{eq:Aexpand}
\frac{A_{m \sigma} (\varepsilon,T,B)}{ A_{m \sigma} (0, 0,0)} 
& = & 1 - \frac{c_\varepsilon \varepsilon^2  + c'_T T^2 + 
c_B B^2 }{\Tk^2} \; ,
\end{eqnarray}
(Particle-hole and spin symmetries forbid terms linear in $\varepsilon$
or $B$.) This implies the same for the resistivity,
\begin{eqnarray}
\label{eq:resistivity-Taylor}
\frac{\rho(T,B)}{\rho(0,0)} & = & 1 - 
\frac{c_T T^2 +  c_B B^2}{\Tk^2} \; ,
\end{eqnarray}
with $c_T = (\pi^2/3)c_\varepsilon + c'_T$.  The so-called
Fermi-liquid coefficients $c_\varepsilon$, $c_T$ and $c_B$ are
universal, $N$-dependent numbers, characteristic of the fully screened
Fermi-liquid fixed point.  For $N=1$, the coefficients $c_T$ and $c_B$
have recently been measured experimentally in transport studies
through quantum dots and compared to theoretical
predictions.\cite{Kretinin2011} The coefficient $c_\varepsilon$ is, in
principle, also measurable via the non-linear conductance of a Kondo
dot coupled strongly to one lead and very weakly to another.\cite{PG}
(The latter condition corresponds to the limit of a
weak tunneling probe; it ensures that the non-linear conductance
probes the \textit{equilibrium} shape of the Kondo resonance, and
hence the equilibrium Fermi-liquid coefficient $c_\varepsilon$.)

The goal of this paper is twofold: first, to analytically establish
the $N$ dependence of $c_\varepsilon$, $c_T$ and $c_B$ using
Fermi-liquid theory similar to NB; and second, to numerically calculate them
using an NRG code that exploits non-Abelian symmetries, in order to
establish a benchmark for the quality of the latter.  Our main results
are as follows: First, if the Kondo temperature is defined by
\begin{eqnarray}
\label{eq:define-Tk}
\Tk  = \frac{N(N+2)}{3\pi \chiimp}  = 
\frac{4 S (S+1)}{3\pi \chiimp} \; , 
\end{eqnarray}
where $\chiimp$ is the static impurity susceptibility at zero
temperature, the Fermi-liquid coefficients are given by 
\begin{eqnarray}
\label{eq:cBTeresults}
c_{B} = \frac{(N+2)^2}{9}  , \quad
c_T = \pi^2 \frac{4N +5}{9}  , \quad 
c_\varepsilon = \frac{2N+7}{6}  . \qph
\end{eqnarray}
For general $N$, the formula for $c_T$ has first been found by
Yoshimori,\cite{Yoshimori} while those for $c_B$ and $c_\varepsilon$
are new (though not difficult to obtain).  Second, our numerical results for
$N=1, 2$ and $3$ are found to agree with the predictions of
\Eq{eq:cBTeresults} to within 5\%.

\section{Fermi-liquid theory}
\label{sec:Fermi-liquid}

In this section, we analytically calculate the Fermi-liquid
coefficients $c_B$, $c_T$ and $c_\varepsilon$ for general $N$. With
the benefit of hindsight, we selectively combine various elements of
the work on FLT of Nozi\`eres,\cite{Nozieres1974}
NB,\cite{Nozieres1980} AL\cite{AL93} and PG\cite{PG}. Detailed
justifications for the underlying assumptions are given by these
authors in their original publications and hence will not be repeated
here. Instead, our goal is to assemble their ideas in such a way that
the route to the desired results is short and sweet.

We begin by summarizing Nozi\`eres' ideas for expressing the $T$ matrix
in terms of scattering phase shifts and expanding the latter in terms
of phenomenological Fermi-liquid parameters. Next, we recount AL's
insight that this expansion can be reproduced systematically by doing
perturbation theory in the leading irrelevant operator of the model's
zero-temperature fixed point. Then we adopt PG's strategy of
performing the expansion in a quasiparticle basis in which the contant
part of the phase shift has already been taken into account, which
considerably simplifies the calculation. Our own calculation is
presented using notation analogous to that of PG, while taking care to
highlight the extra terms that arise for $N>1$. It turns out that
their extra contributions can be found with very little extra effort.

\subsection{Phase shift and $T$ matrix}
Since the ground state of the fully screened Kondo model is a spin
singlet, a low-energy quasiparticle scattering off the impurity
experiences strong elastic scattering as if the impurity were
nonmagnetic. Moreover, it also experiences a weak local interaction
if some energy $(\ll \Tk)$ is available to weakly excite the singlet,
causing some inelastic scattering.  Since the singlet binding energy
is $\Tk$, the strength of this local interaction is
proportional to $1/\Tk$.

Nozi\`eres\cite{Nozieres1974} realized that this combination of strong
elastic scattering and a weak local interaction can naturally be
treated in terms of scattering phase shifts.  The phase shift of a
quasiparticle with quantum numbers $m\sigma$ and excitation energy
$\varepsilon$ relative to the Fermi energy can be written as
\begin{eqnarray}
\label{eq:tildedelta}
\delta_{m\sigma} (\varepsilon) = \delta_{m\sigma}^0 + 
\tdelta_{m \sigma} (\varepsilon) \; , 
\quad \delta_{m\sigma}^0 = \pi/2 \; .
\end{eqnarray}
Here $\delta_{m\sigma}^0$ is  the phase shift for 
$\varepsilon = B=T=0$; it has the maximum
possible value for scattering off a non-magnetic impurity, namely
$\pi/2$.  Finite-energy corrections 
arising from weak excitations of the singlet 
are described by $\tdelta_{m \sigma} (\varepsilon)$, which 
is proportional to $1/\Tk$. 

If inelastic scattering is weak, unitarity of the $S$ matrix can be
exploited\cite{Nozieres1974} to write the $T$ matrix 
in the following form
(we use  the notation PG\cite{PG};
for a detailed analysis, see AL's discussion\cite{AL93}
of the terms arising from their Figs.~6 and 7):
 \begin{eqnarray}
\label{eq:define-phase-shift-Tmatrix-inelastic}
1 - 2 \pi \nu i \T_{m \sigma}(\varepsilon) & =  &  e^{2 i \delta_{m\sigma}(\varepsilon)} 
\bigl[1 - 2 \pi \nu i \tTin_{m\sigma} (\varepsilon)
\bigr]
\; .  \qquad 
\end{eqnarray}
Here $\tTin$ accounts for weak inelastic two-body scattering
processes, and is proportional to $1/\Tk^2$. It is to be
calculated in a basis of quasiparticle states in which the phase shift
$\delta^0_{m\sigma}$ has already been accounted for.  (Here and below,
tildes will be used on quantities defined with respect to the new basis if
they differ from corresponding ones in the original basis.)

Expanding \Eq{eq:define-phase-shift-Tmatrix-inelastic} in the small
(real) number $\tdelta_{m\sigma}(\varepsilon)$ and recalling that
$e^{2 i \delta^0_{m \sigma}} = -1$, one finds that the imaginary part
of the $T$ matrix, which determines the spectral function, can be
expressed as
\begin{eqnarray}
\label{eq:ImToriginal}
 - \pi \nu \textrm{Im} \T_{m\sigma} ( \varepsilon)
= 1 - \bigl[ \tdelta^2_{m \sigma}(\varepsilon) - \pi \nu 
\textrm{Im} \tTin_{m\sigma} (\varepsilon) \bigr] \; ,  \qqph
\end{eqnarray}
to order $1/\Tk^2$. Comparing this to \Eq{eq:Aexpand}, we conclude
that knowing $\tdelta$ to order $1/\Tk$ and $ \tTin$ to order
$1/\Tk^2$ suffices to fully determine the Fermi-liquid coefficients
$c_B$, $c_T$ and $c_\varepsilon$.

Now, a systematic calculation of $\tdelta$ and $\tTin$ requires a
detailed theory for the strong-coupling fixed point, which became
available only with the work of AL in the early 1990s.  Nevertheless,
Nozi\`eres succeeded in treating the case $N=1$ already in
1974,\cite{Nozieres1974} using a phenomenological expansion of
$\tdelta_{m\sigma} (\varepsilon)$ in powers of
$(\varepsilon-\Ezsigma)/\Tk$ [$\Ezsigma$ represents the Zeeman energy
of quasiparticles in a magnetic field, see \Eq{eq:HPG}
below] and $\barn_{m'\sigma'} = n_{m'\sigma'} - n^0_{m'\sigma'}$,
the deviation of the total quasiparticle number $n_{m'\sigma'}$ from
its ground-state value. The prefactors in this expansion have the
status of phenomenological Fermi-liquid parameters. Using various
ingenious heuristic arguments, he was able to show that all these
parameters, and also $\tTin$, are related to each other and can be
expressed in terms of a single energy scale, namely the Kondo
temperature.  Moreover, by choosing the prefactor of $\varepsilon$ in
this expansion to be $1/\Tk$, he suggested a definition of the Kondo
temperature that also fixes its numerical prefactor. (Our paper adopts
this definition, too.)  In 1980, NB generalized this strategy
\cite{Nozieres1980} to general $N$, finding an expansion of the form
\begin{eqnarray}
\nonumber
\tdelta_{m\sigma} (\varepsilon)
& = &  \alpha (\varepsilon - \Ezsigma) - 3 \psi \barn_{m, - \sigma} 
\\
\label{eq:NB-phase-expansion}
& & {} + \psi \sum_{m' \neq m} 
(\barn_{m' \sigma} - \barn_{m', - \sigma } )\; , 
\end{eqnarray}
where $\alpha$ and $\psi$ are phenomenological Fermi-liquid parameters
related by $\alpha = 3 \psi \nu=1/\Tk$. [NB's initial 
version of \Eq{eq:NB-phase-expansion} [their Eq.~(34)] does
not contain the Zeeman contribution $\Ezsigma$, but the latter
is implicit in their subsequent treatment of the 
Zeeman field before their Eq.~(37).]

In the following subsections, we show how NB's expansion for $\tdelta$
can be derived systematically.  AL\cite{AL93} and PG\cite{PG} have
shown how to do this for $N=1$; we will generalize their discussion to
arbirtrary $N$.

\subsection{Leading irrelevant operator}
\label{sec:leading-irrelevant}

AL showed\cite{AL93} that NB's heuristic results can be derived in a
systematic fashion by doing perturbation theory in the leading
irrelevant operator of the model's zero-temperature fixed point.  As
perturbation, they took the operator with the lowest scaling dimension
satisfying the requirements of being (i) local, (ii) independent of
the impurity spin operator $\vec S$, since the latter is fully
screened, (iii) SU(2)-spin-invariant, (iv) and independent of the
local charge density, just as the Kondo interaction. The
operator sastifying these criteria has the form\cite{A90}
\begin{eqnarray}
\label{eq:AL-start-JJ}
H_\lambda = - \lambda  :\! \vec J (0) \cdot \vec J (0)\! :  \; ,
\end{eqnarray}
where $\vec J(0)$ is the quasiparticle spin density at the impurity
site, and $: \! \dots \! :$ denotes the point-splitting regularization
procedure (see Appendix).  In Appendix~D of
Ref.~\onlinecite{AL93}, AL showed in great detail how NB's phase
shifts can be computed using \Eq{eq:AL-start-JJ}, for the
single-channel case of $N=1$.  They did not devote as much attention
to the case of general $N$, though the needed generalizations are
clearly implied in their work. We here present the corresponding
calculation in some detail, following the notational conventions of
PG, which differ from those of AL in some regards (see
Appendix). The main difference is that PG formulate
the perturbation expansion in a new basis of
quasiparticle states,  in which the phase shift $\delta^0_{m\sigma}$ has
already been accounted for, which somewhat simplifies the discussion.
(We remark that PG chose $\delta^0_{m\sigma} = \sigma \pi/2$ rather
than $\pi/2$ as used by NB and us, but the extra $\sigma$ has no
consequences for the ensuing arguments.)

\begin{center} 
\begin{figure}[t]
\includegraphics[angle=0,width=\columnwidth]{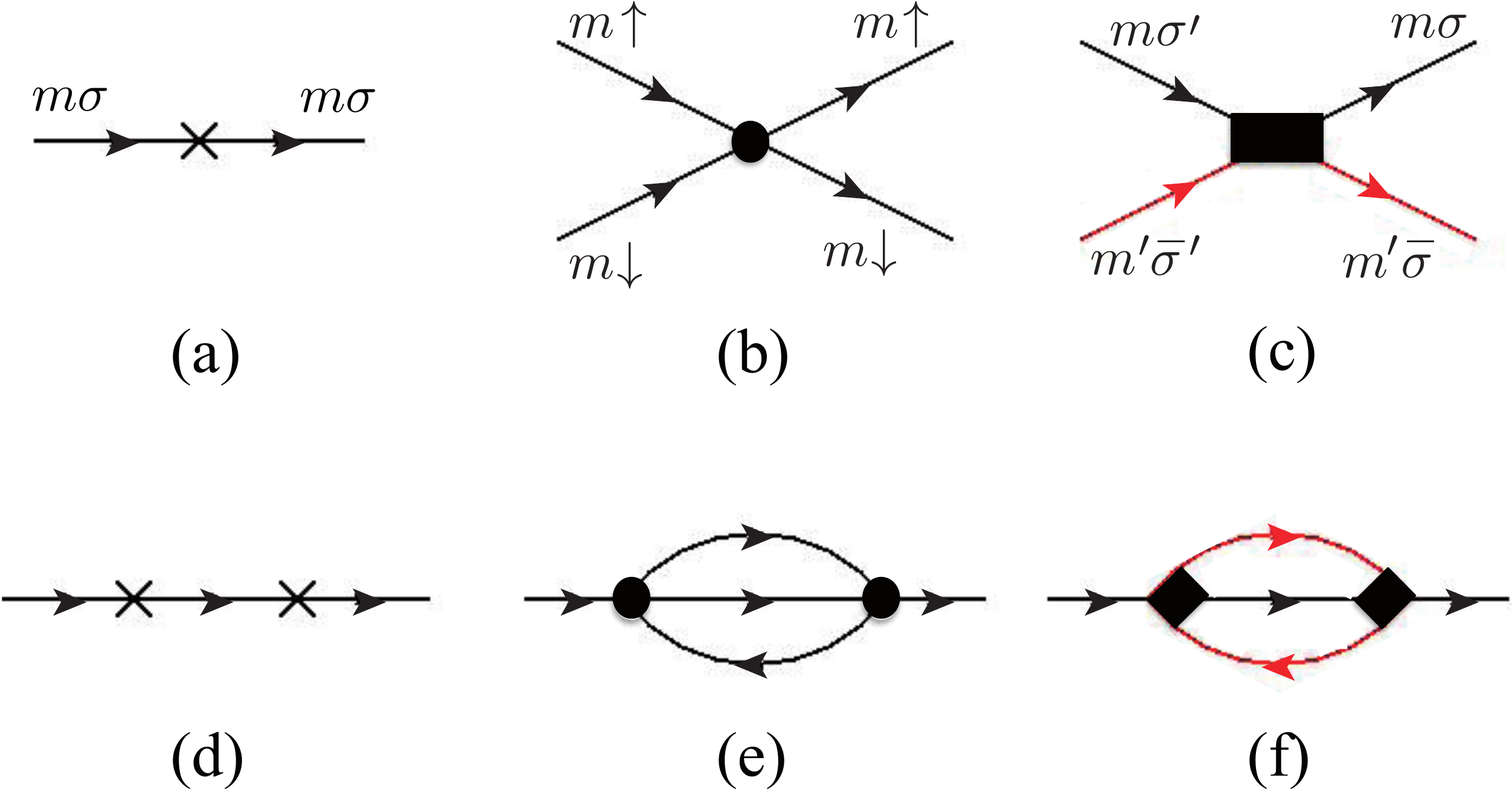}
\caption{(Color online) (a)-(c) Vertices associated with 
$\Hone$, $\Htwo$ and $\Hthree$, respectively. 
(d)-(f) Nonzero second-order contributions to 
the quasiparticle self-energy, $\tSigma_{m\sigma}$,
involving $\Hone^2$, $\Htwo^2$ and $\Hthree^2$, respectively.
The  contributions involving $\Hone \Htwo$, $\Hone \Hthree$
and $\Htwo \Hthree$ all vanish, the former two due to the odd power of
energy in the two-leg vertex.
}
\label{fig:one}
\end{figure}
\end{center}

The quasiparticle Hamiltonian describing the vicinity of the
strong-coupling fixed point (fp) has the form
\begin{eqnarray}
\Hfp = H_{\rm fp,0} + H_\lambda \; , 
\end{eqnarray}
where
\begin{eqnarray}
\label{eq:HPG}
H_{\rm fp,0}  =  \sum_{m\sigma k} (\xi_k + \Ezsigma) \! : \! 
\psi^\dagger_{km\sigma}   \psi_{km\sigma} \!  : \; , \quad
\Ezsigma = - \frac{\sigma B}{2}  \qqph
\end{eqnarray}
describes free quasiparticles in a magnetic field $B$, with Zeeman
energy $\Ezsigma$. Note that although the Zeeman term in the bare
Hamiltonian (\ref{eq:Kondomodel}) is local, it is global in
\Eq{eq:HPG}, because the effective quasiparticle Hamiltonian $\Hfp$
contains no local spin.  Using standard point-splitting techniques,
which we review in pedagogical detail in the Appendix, the
leading irrelevant operator (\ref{eq:AL-start-JJ}) can be written as
$H_\lambda = \Hone + \Htwo + \Hthree$, with
\begin{subequations}
\label{eq:H-momentum-main}
\begin{eqnarray}
\label{eq:Hel-momentum-main}
\Hone & = & - \frac{1}{2 \pi \nu \Tk}
\sum_{m\sigma kk'} (\xi_k + \xi_{k'}) 
: \! \psi^\dagger_{km\sigma}   \psi_{k'm\sigma} \! : \; , \quad
\phantom{.}
\\
  \label{eq:Htwo-momentum}
\Htwo &= &  \frac{1}{\pi \nu^2 \Tk} \sum_m 
:\! \rho_{m\uparrow} \rho_{m \downarrow}\! : \; , 
\\
  \label{eq:Hthree-momentum}
\Hthree & = & - \frac{2}{3 \pi \nu^2 \Tk}
\sum_{m \neq m'} : \! \vec j_m \cdot \vec j_{m'} \! : \; , 
\end{eqnarray}
\end{subequations}
where 
\begin{subequations}
\begin{eqnarray}
\label{eq:Hint-momentum-main}
 \rho_{m\sigma} & = &
\sum_{kk'\sigma} \psi^\dagger_{km\sigma}   \psi_{k'm\sigma} \; , 
\\
\vec j_m   &=&   \frac{1}{2}
\sum_{kk'\sigma \sigma'} \psi^\dagger_{km\sigma} \vec \tau_{\sigma\sigma'}
  \psi_{k'm\sigma'}  \; . \qph
\end{eqnarray}
\end{subequations}
Here we have expressed the coupling constant $\lambda$
in terms of the inverse Kondo temperature using 
[cf.\ \Eq{eq:identify-alpha1-phi1}]
\begin{eqnarray}
\label{eq:identify-alpha-Tk}
\lambda = \frac{8 \pi (\hbar \vF)^2}{3\Tk}  \; , 
\end{eqnarray}
with the numerical proportionality factor chosen such that $\Tk$
agrees with definition of the Kondo temperature used by NB and PG, as
discussed below. Importantly, the point-splitting procedure fixes the
relative prefactors arising in $\Hone$, $\Htwo$ and $\Hthree$ (whereas
NB's approach requires heuristic arguments to fix them).  Our notation
for $\Hone$ and $\Htwo$ coincides with that used by PG. $\Hthree$
contains all new contributions that enter additionally for $N>
1$. Figure~\ref{fig:one} gives a diagrammtic depiction of all three
terms.

\subsection{First order terms}
\label{sec:phase-shift}

Our first goal is to recover NB's expansion of the phase shift
$\tdelta$ to leading order in $\varepsilon- \Ezsigma$ and $\barn$.
Following PG, this can be done by calculating $\tdelta$ perturbatively
to first order order in $1/\Tk$, in the new basis of quasiparticle
states that already incorporate the phase shift $\delta^0$.  To order
$1/\Tk$, no inelastic scattering occurs, and $\tdelta$ 
is related to the elastic $T$ matrix by
\begin{eqnarray}
\label{eq:define-phase-shift-Tmatrix}
  e^{2 i \tdelta_{m\sigma}(\varepsilon)} = 
1 - 2 \pi \nu i \tTel_{m \sigma}  (\varepsilon) \; .
\end{eqnarray}
The elastic $T$ matrix, in turn, equals the real part of the
quasiparticle self-energy, $\tTel_{m \sigma} (\varepsilon) =
\textrm{Re} \tSigma_{m\sigma} (\varepsilon)$.  (Actually, to order
$1/\Tk$, the self-energy is purely real.)  By expanding
\Eq{eq:define-phase-shift-Tmatrix} for small $\tdelta$, the phase
shift is thus seen to be given by the real part of the self-energy:
\begin{eqnarray}
  \label{eq:expandphaseshift-T-matrix}
  \tdelta_{m\sigma} (\varepsilon) \simeq - \pi \nu 
\textrm{Re} \tSigma_{m \sigma} (\varepsilon) \; . 
\end{eqnarray}

Now, as pointed out already by Nozi\`eres in 1974,\cite{Nozieres1974}
a first-order perturbation calculation of the self-energy  is equivalent
to treating interaction terms in the mean-field (MF) approximation. They 
then take the form 
\begin{subequations}
\begin{eqnarray}
\label{eq:HartreeFock}
\HtwoMF \! &= &  \frac{1}{\pi \nu^2 \Tk} \sum_{m\sigma}
:\! \rho_{m\sigma} \! \!: \barn_{m , - \sigma  } 
\\
\HthreeMF \! & = & - \frac{1}{3 \pi \nu^2 \Tk} \!\!
\sum_\sigma \!\! \! \sum_{m \neq m'} \!\! \! \! : \! \rho_{m\sigma } \! : 
\!(\barn_{m' \sigma } - \barn_{m', - \sigma } ) , \qqph
\end{eqnarray}
\end{subequations}
where $\barn_{m\sigma} = \langle :\! \rho_{m\sigma} \! : \rangle$,
the quasiparticle number  relative to the $B=0$ ground state, 
is given by 
\begin{eqnarray}
  \label{eq:quasiparticle-density-in-field}
  \barn_{m\sigma} = - \nu \Ezsigma =  \sigma \nu B/2 \; \; . 
\end{eqnarray}
The mean-field version of the 
leading irrelevant operator thus has the form
\begin{eqnarray}
  \label{eq:Hsingle-particle}
H_\lambda^{\rm MF} 
& = & 
\sum_{m\sigma kk'} h_{m\sigma} (\xi_k , \xi_{k'})
: \! \psi^\dagger_{km\sigma}   \psi_{k'm\sigma} \! : \; , \qqph
\\
h_{m\sigma} (\xi_k , \xi_{k'}) & = & 
 \frac{1}{\pi \nu \Tk} \Biggl[ - \frac{1}{2}(\xi_k + \xi_{k'}) 
+ \frac{\barn_{m, - \sigma}}{\nu} \Biggr. \\
& &
\nonumber  \quad \qquad \Biggl.  -  \sum_{m' \neq m} 
\frac{ \barn_{m' \sigma} - \barn_{m', -\sigma}}{3 \nu}
 \Biggr] \; . 
\end{eqnarray}
For such a single-particle perturbation, the self-energy can be
directly read off from $h_{m\sigma}$ using
\begin{eqnarray}
  \label{eq:read-off-sigma-from-h}
  \tSigma_{m\sigma}(\varepsilon) = h_{m\sigma}
(\varepsilon - \Ezsigma, \varepsilon - \Ezsigma) \; , 
\end{eqnarray}
because $k$ sums of the type  $\sum_k 1/(\varepsilon - \xi_k -
\Ezsigma+ i0^+)$  yield residues involving
$\xi_{k} = \varepsilon - \Ezsigma$.
Using \Eq{eq:read-off-sigma-from-h} in
\Eq{eq:expandphaseshift-T-matrix} for the phase shift, we find
\begin{eqnarray}
  \label{eq:GP-phase-shift-a-la-NB}
  \tdelta_{m\sigma}(\varepsilon) & = &
 \frac{1}{\Tk} \Biggl[\varepsilon - \Ezsigma
- \frac{\barn_{m, - \sigma}}{\nu} 
\Biggr. \\
& &
\nonumber  \quad \qquad \Biggl.  
+  \sum_{m' \neq m} 
\frac{ \barn_{m' \sigma} - \barn_{m', -\sigma}}{3\nu} \Biggr] \; .
\end{eqnarray}
This fully agrees with the expansion (\ref{eq:NB-phase-expansion})
of NB if we make the identification $1/\Tk = \alpha = 3 \psi \nu$,
thus confirming the validity of NB's heuristic arguments.
Note that the coefficient of $\varepsilon- \Ezsigma$ 
in \Eq{eq:GP-phase-shift-a-la-NB} comes
out as $1/\Tk$, in agreement with the conventions
of NB and PG, as intended by our choice
of numerical prefactor in \Eq{eq:identify-alpha-Tk}.
 
As consistency check, let us review how NB used
\Eq{eq:GP-phase-shift-a-la-NB} to calculate the Wilson ratio. First,
\Eq{eq:GP-phase-shift-a-la-NB} implies an impurity-induced change in
the density of states per spin and channel of $\nu^\imp_{m\sigma}
(\varepsilon) = \frac{1}{\pi} \partial_\varepsilon
\delta_{m\sigma}(\varepsilon)$. This yields a corresponding
impurity-induced change in the specific heat, $\Cimp$.  At zero field
(where $\Ezsigma$ and $\barn_{m\sigma}$ vanish), the change
relative to the bulk is given by
\begin{eqnarray}
  \label{eq:specific-heat}
  \frac{\Cimp}{C} = \frac{2 N \nu^\imp_{m\sigma} (0) }{2N \nu}
  = \frac{1}{\pi \nu \Tk} \; . 
\end{eqnarray}
Second, the Friedel sum rule 
for the impurity-induced change in local charge in channel $m$ for spin $\sigma$
at $T=0$ gives 
\begin{eqnarray}
  \label{eq:Friedel-charge}
  N^\imp_{m\sigma}  & = &  \frac{1}{\pi} \delta_{m\sigma}(0) = 
  \frac{1}{2} + \frac{1}{\pi} \tdelta_{m\sigma}(0) \; , 
\end{eqnarray}
and \Eq{eq:GP-phase-shift-a-la-NB}, together
with \Eq{eq:quasiparticle-density-in-field} for $\barn_{m \sigma}$, 
leads to 
\begin{eqnarray}
\label{eq:tdeltaB}
\tdelta_{m\sigma}(0) 
& = &  \frac{\sigma B}{\Tk} \left[ \frac{1}{2} + \frac{1}{2} +
\frac{N-1}{3} \right] 
=   \frac{\sigma B (N+2)}{3 \Tk} \; . \qqph
\end{eqnarray}
The linear response of the impurity-induced magnetization,
 $M^\imp =
\frac{1}{2}\sum_m(N^\imp_{m\uparrow} - N^\imp_{m\downarrow})$,
then gives the impurity contribution to the spin susceptibility as
\begin{eqnarray}
  \label{eq:Friedel-magnetization}
  \chiimp = \frac{M^\imp}{B}  = 
\frac{N(N+2)}{3\pi \Tk} = 
\frac{4 S(S+1)}{3\pi \Tk}  \; . 
\end{eqnarray}
(For all expressions involving $\chiimp$ here and below, the
limit $B\to 0$ is implied.) The corresponding bulk contribution is
$\chi = \nu N /2$.  Thus, the Wilson ratio is found to be
\begin{eqnarray}
  \label{eq:Wilson-define}
  R = \frac{\chiimp/\chi}{\Cimp/C} 
= \frac{2(N+2)}{3} =  \frac{4(S+1)}{3} \; , 
\end{eqnarray}
in agreement with more elaborate calculations by
Yoshimori\cite{Yoshimori} and by Mih\'aly and Zawadowski.\cite{Zaw}

Note that \Eq{eq:Friedel-magnetization} relates Nozi\`eres' definition
of the Kondo temperature to an observable quantity, $\chiimp$, that
can be calculated numerically.  We used this as a precise way of
defining $\Tk$ in our numerical work.  (Subtleties
involved in calculating $\chiimp$ are discussed 
in Sec.~\ref{sec:define-TK}.) Note that
up to a prefactor, \Eq{eq:Friedel-magnetization}
for $\chiimp$ has the form $\chifree(\Tk)$,
where $\chifree (T) = S(S+1)/(3 T)$ is
the static susceptibility of a free spin $S$ at temperature $T$.

We are now in a position to extract our first Fermi-liquid
coefficient, $c_B$. For this, it suffices to know the spectral
function $A$ in \Eq{eq:Aexpand} to quadratic order in $B$, at
$\varepsilon =T=0$, where $\tTin=0$.  Inserting the corresponding
expression (\ref{eq:tdeltaB}) for $\tdelta_{m\sigma}(0)$ into
\Eq{eq:ImToriginal} for $\textrm{Im} \T$, we find
\begin{eqnarray}
  \label{eq:A-B-dependence}
  A_{m\sigma}(0,0,B) = \frac{1}{\nu \pi^2} 
\left[ 1 -   \frac{(N+2)^2}{9} \frac{B^2}{\Tk^2} \right] \; . 
\end{eqnarray}
Comparing this to \Eq{eq:Aexpand}, we read off $c_B = (N+2)^2/9$.

Note that if the definition (\ref{eq:Friedel-magnetization}) 
of $\Tk$ in terms of $\chiimp$ is taken as given, 
$c_B$ can actually be derived on the back of an envelope: for a
fully screened Kondo model, the impurity-induced spin susceptibility
gets equal contributions from all $N$ channels, $\chiimp = N
\chiimp_{m}$, and the Friedel sum rule relates the contribution
from each channel to phase shifts, $\chiimp_{m} = M^\imp_m/B =
[\tdelta_{m\uparrow}(0) - \tdelta_{m\downarrow}(0)]/(2 \pi B)$, implying
$\tdelta_{m\sigma} (0)= \sigma (\pi \chiimp/N)B$. Using this in
\Eq{eq:ImToriginal} yields
\begin{eqnarray}
  \label{eq:A-B-dependence-alternative}
  A_{m\sigma}(0,0,B) = \frac{1}{\nu \pi^2} 
\Bigl[ 1 -   (\pi \chiimp/N)^2B^2 \Bigr] \; , 
\end{eqnarray}
which is equivalent to \Eq{eq:A-B-dependence} if
\Eq{eq:Friedel-magnetization} holds.  

\subsection{Second order terms}
We next discuss inelastic scattering for $B=0$, but at finite
temperature. To order $1/\Tk^2$, inelastic scattering is described by
the imaginary part of the quasiparticle self-energy arising from the
second-order contributions of $\Hone$, $\Htwo$ and $\Hthree$, shown in
diagrams (d)-(f) of Fig.~\ref{fig:one}, respectively. These
diagrams give
\begin{subequations}
  \label{eq:Tinelastic123}
\begin{eqnarray}
  {\rm Im} \tSigmaone_{m\sigma}(\varepsilon)
 & = & -  \frac{\varepsilon^2}{\pi \nu \Tk^2} \; , 
\\
  {\rm Im} \tSigmatwo_{m\sigma}(\varepsilon)
 & = & -  \frac{\varepsilon^2 + \pi^2 T^2}{2 \pi \nu  \Tk^2} \; , 
\\
  {\rm Im} \tSigmathree_{m\sigma}(\varepsilon)
 & = &  \frac{2}{3} (N-1) \, {\rm Im} \tSigmatwo_{m\sigma}(\varepsilon) \; .  
\end{eqnarray}
\end{subequations}
The first two can also be found in the discussion of PG, whose
strategy we follow here.  (They also appear, in slightly different
guise, in the discussion of AL\cite{AL93}.)  The third is proportional
to the second, and the factor $2/3$ originates from $(2/3)^22 s(s+1) $
with $s=1/2$, since the relative prefactor between $\Hthree$ and
$\Htwo$ brings in two powers of $2/3$, and the algebra of Pauli
matrices yields a factor $2s(s+1)$.

Now, the term called $\tTin$ in
\Eq{eq:define-phase-shift-Tmatrix-inelastic} 
by definition describes the 
contribution of the \textit{two-body} terms
$\Htwo$ and $\Hthree$ to inelastic scattering:
\begin{eqnarray}
  \label{eq:inelastic-tT-matrix}
\textrm{Im} \tTin_{m\sigma}(\varepsilon) =  \textrm{Im} \Bigl[
\tSigmatwo_{m\sigma}(\varepsilon) +
\tSigmathree_{m\sigma}(\varepsilon) \Bigr] \; .
\end{eqnarray}
The contribution ${\rm Im} \tSigmaone$ from $\Hone$ is \textit{not}
included in $\textrm{Im} \tTin$ here, since it actually equals $-\tdelta^2/ \pi
\nu$, and hence is already contained in the factor $e^{2 i \tdelta}$
in \Eq{eq:define-phase-shift-Tmatrix-inelastic}.  Indeed,
in \Eq{eq:ImToriginal} for
the imaginary part of the $T$ matrix in the original basis,
the $\tdelta^2$ term equals $- \pi \nu {\rm Im}\tSigmaone$.
Collecting all ingredients, \Eq{eq:ImToriginal} 
gives 
\begin{eqnarray}
\nonumber
\lefteqn{A_{m\sigma}(\varepsilon,T,0) }
\\
\label{eq:ImTinTotal}
& = & \frac{1}{\nu \pi^2} \left[ 
1  - \frac{\varepsilon^2}{\Tk^2} - 
\frac{\varepsilon^2 + \pi^2 T^2}{2 \Tk^2} \Bigl(1 + \frac{2}{3}(N-1) 
\Bigr) \right] 
\nonumber
\\
  \label{eq:AeT-original}
& = & 
\frac{1}{\nu \pi^2} \left[ 
1 - \frac{(2N+7)\varepsilon^2 + (2N+1)\pi^2 T^2}{6 \Tk^2}
\right] \; .  
\qqph
\end{eqnarray}
For $N=1$, the second term reduces to the familiar form $- (3
\varepsilon^2 + \pi^2 T^2)/(2 \Tk^2)$ found by AL\cite{AL93} and
GP\cite{PG}.  Comparing \Eqs{eq:AeT-original} and 
(\ref{eq:Aexpand}) and (\ref{eq:resistivity-Taylor})
we read off $c_\varepsilon = (2N+7)/6$ and $c'_T = \pi^2 (2N+1)/6$,
implying $c_T = \pi^2 (4N+5)/9$.

\section{NRG results}
\label{sec:NRG}

In this section, we describe our NRG work.  We had set ourselves the
goal of achieving an accuracy of better than 5\% for the Fermi-liquid
coefficients. To achieve this, two ingredients were essential. First,
exploiting non-Abelian symmetries; and second, defining the Kondo
temperature with due care. The latter is a matter of some
subtlety \cite{Hanl2013a} because the wide-band limit assumed in
analytical work does not apply in numerical calculations.

We begin below by giving the Lehmann representation for the desired
spectral function. We then discuss the non-Abelian symmetries used in
our NRG calculations and explain how the Kondo temperature was
extracted numerically.  Finally, we present our numerical results.

\subsection{NRG details}
\label{sec:NRG-details}

To numerically calculate the $T$ matrix of \Eq{eq:define-T-matrix},
we use equations of
motion\cite{Costi2000,Rosch2003} to express it as
\begin{subequations}
\begin{eqnarray}
{\cal T}_{m\sigma}(\varepsilon) &= & J_{\rm K}\langle S_z \rangle +  
\langle\langle O_{m\sigma}; O^{\dag}_{m\sigma} \rangle\rangle,
\\
O_{m\sigma} & \equiv & [\Psi_{m\sigma}(0),\Hloc] =
J_{\rm
  K}\sum_{\sigma'}{\vec S} \cdot \frac{{\vec \tau}_{\sigma \sigma'}}{2}
\Psi_{m\sigma'}(0) . \qqph
\end{eqnarray}
\end{subequations}
Here $\langle\langle\,\cdot\,; \cdot\,
\rangle\rangle$ denotes a retarded correlation function, and
$\Psi_{m\sigma}(0)=\frac{1}{\sqrt{N_\discrete}}\sum_k c_{km\sigma}$,
where $N_\discrete$ is the number of discrete levels in the
band (and hence proportional to the system size). The
spectral function is then calculated in its Lehmann-representation,
\begin{eqnarray}
\lefteqn{
A_{m\sigma}(\varepsilon,T,B)  =}
\nonumber
\\ 
& \quad \sum_{a,b}\frac{e^{-\beta E_a}+e^{-\beta E_b}}{Z}\vert \langle a \vert O_{m\sigma} \vert b \rangle \vert^2 \delta (\varepsilon - E_{ab}), 
\label{eq:discdata}
\end{eqnarray}
with $E_{ab}=E_b-E_a$, using the full density matrix (FDM) approach of 
NRG.\cite{Weichselbaum2007,Anders2005,Peters2006,Wb2012a}

For our numerical work, we take the conduction band energies to lie
within the interval $\xi_k \in [-D,D]$, with Fermi energy at 0 and
half-bandwidth $D=1$, and take the density of states per spin, channel
and unit length to be constant, as $1/2D$.  (It is related to the
extensive density of states used in Sec.~\ref{sec:Fermi-liquid} by
$\nu = N_\discrete/2D$.)  For the calculations used to determine the
Fermi-liquid parameters, we use exchange coupling $\nu \Jk =0.1$, so
that the Kondo temperature $T_{\rm K}/D \propto \exp[-1/(\nu J_{\rm
K})]$ has the same order of magnitude for $N=1, 2$ and 3, namely
$\lesssim 10^{-4}$. Following standard NRG
protocol,\cite{Wilson1975,Krishnamurthy1980,Bulla2008} the conduction
band is discretized logarithmically with discretization parameter
$\Lambda$, mapped onto a
Wilson chain, and diagonalized iteratively.  NRG truncation at each
iteration step is controlled by either specifying the number of kept
states per shell, $N_{\rm K}$, or the truncation energy, $E_{\rm tr}$
(in rescaled units, as defined in Ref.~\onlinecite{Weichselbaum2011}),
corresponding to the highest kept energy per shell. Spectral data are
averaged over $N_z$ different, interleaving logarithmic discretization
meshes.\cite{Oliveira1991} The values for NRG-specific parameters used
here are given in legends in the figures below.

For the fully screened $N$-channel Kondo model, the dimension of the
local Hilbert space of each supersite of the Wilson chain is
$4^N$. Since this increases exponentially with the number of channels,
it is essential, specifically so for $N=3$, to reduce
computational costs by exploiting non-Abelian symmetries
\cite{Weichselbaum2012b} to combine degenerate states into
multiplets. Several large symmetries are available\cite{Pang1992}: For
$B=0$, the model has SU(2)$\times$U(1)$\times$SU$(N)$
spin-charge-channel symmetry.  If the bands desribed by $H_0$ are
particle-hole symmetric, as assumed here, the model also has a
SU(2)$\times$[SU(2)]${}^N$ spin-(charge)${}^N$ symmetry, involving
SU(2) mixing of particles and holes in each of the $N$ channels. The
U(1)$\times$SU$(N)$ and [SU(2)]${}^N$ symmetries are not mutually
compatible (their generators do not all commute), however,
implying that both are subgroups of a larger symmetry group, the
symplectic Sp$(2N)$. Thus the full symmetry of the model for $B=0$ is
SU(2)$\times$Sp$(2N)$.  For $B\neq 0$ it is U(1)$\times$Sp$(2N)$,
since a finite magnetic field breaks the SU(2) spin symmetry to the
Abelian U(1) $S_z$ symmetry.  When the model's \textit{full} symmetry
is exploited, the multiplet spaces encountered in NRG calculations
exhibit \textit{no} more degeneracies in energy at all.

Using only Abelian symmetries turned out to be clearly
insufficient to obtain well converged numerical data for $N=3$,
despite having a relatively large $\Lambda$. This, however, is required
for accurate Fermi-liquid coefficients with errors below a few
percent. For numerically converged data, therefore, it was essential
to use non-Abelian symmetries. For our $B=0$ calculations, it
turned out to be sufficient to use SU(2)$\times$U(1)$\times$SU$(N)$
symmetry for calculating $c_T$, but the full SU(2)$\times$Sp$(2N)$
symmetry was needed for calculating $c_\varepsilon$. Likewise, for our
$B\neq 0$ calculations of $c_B$, we needed to use the full
U(1)$\times$Sp$(2N)$ symmetry. Doing so led to an enormous
reduction in memory requirements, the more so the larger the rank of
the symmetry group [Sp$(2N)$ has rank $N$, and SU$(N)$ has rank
$N-1$].
For $N=3$, for example, we
kept $\lesssim 13\,500$ multiplets for
SU(2)$\times$U(1)$\times$SU$(3)$ or $\lesssim 3\,357$ multiplets for
SU(2)$\times$Sp$(6)$ during NRG truncation, which, in effect,
amounts to keeping $\lesssim 980\,000$ individual
states.\cite{Weichselbaum2012b}

\subsection{Definition of $\Tk$}
\label{sec:define-TK}

The Fermi-liquid theory of Sec.~\ref{sec:Fermi-liquid} implicitly
assumes that the model is considered in the so-called scaling limit,
in which the ratio of Kondo temperature to bandwidth vanishes, $\Tk/D
\to 0$.  In this limit, physical quantities such as
$\rho(T,B)/\rho(0,0)$ 
are universal scaling functions, which
depend on their arguments only in the combinations $B/\Tk$ and
$T/\Tk$.  Since the shape of such a scaling function, say
$\rho(0,B)/\rho(0,0)$ plotted versus $B/\Tk$, 
is universal, i.e.\ independent of the bare parameters (coupling $\Jk$
and bandwidth $D$) used to calculate it, curves generated by
different combinations of bare parameters can all be made to collapse
onto each other by suitably adjusting the parameter $\Tk$ for each.
In the same sense the Fermi-liquid parameters $c_B$, $c_T$ and
$c_\varepsilon$, being Taylor-coefficients of universal curves, are
universal, too.

One common way to achieve a scaling collapse, popular
particularly in experimental studies, is to identify the Kondo
temperature with the field $B_\half$ or temperature $T_\half$ at
which the impurity contribution to the resisitivity has decreased to
half its unitary value,
\begin{eqnarray}
  \label{eq:ThalfBhalf}
\rho(0,B_\half)= \rho(0,0)/2 \; , 
\quad
\rho(T_\half,0) =    \rho(0,0)/2  \; . \qqph 
\end{eqnarray}
However, this is approach is not suitable for the purpose of
extracting Fermi-liquid coefficients, for which $\Tk$ has to be
defined in terms of (analytically accessible) low-energy properties
characteristic of the strong-coupling fixed point.  In
Sec.~\ref{sec:Fermi-liquid} we have therefore adopted Nozi\`eres'
definition of $\Tk$ in terms of the leading energy dependence of the
phase shift $\tdelta_{m\sigma}^0$ [\Eq{eq:GP-phase-shift-a-la-NB}],
implying that it can be expressed in terms of $\chiimp$, of the local
static spin susceptibility at zero temperature
[\Eq{eq:Friedel-magnetization}].  In the scaling limit, this
definition of $\Tk$ matches $B_\half$ or $T_\half$ up to prefactors,
i.e.\ $B_\half/\Tk$ and $T_\half/\Tk$ are universal, $N$-dependent
numerical constants, independent of the model's bare parameters.

\begin{figure}
\includegraphics{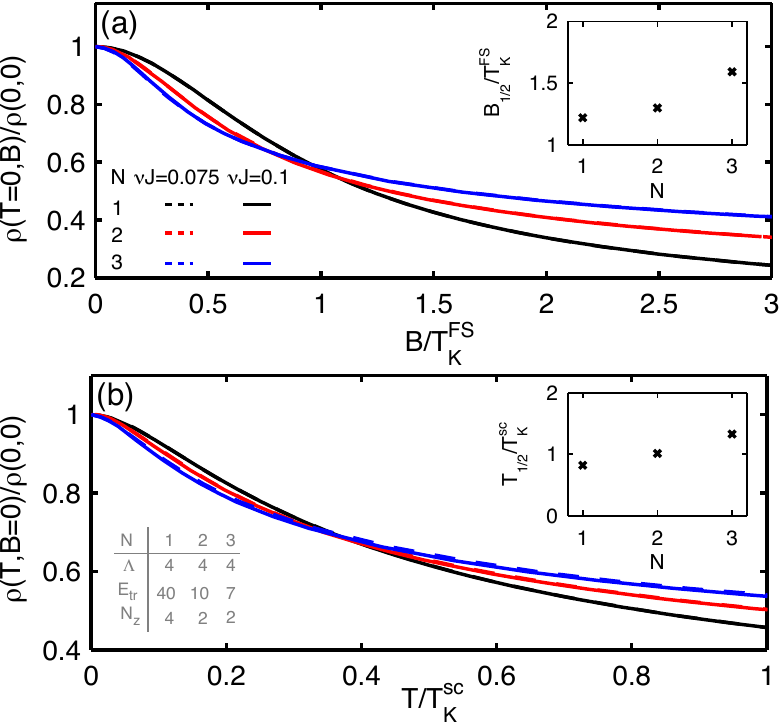}
\caption{(Color online) Scaling collapse of (a) the resistivity at
  zero temperature as function of field, and (b) at zero field as
  function of temperature, calculated for two different values of the
  bare coupling, $\nu \Jk$ (dashed or solid), and for $N=1, 2$ and
  $3$.  For each $N$, the dashed and solid curves overlap so well that
  they are almost indistinguishable.  The insets compare the energy
  scales $B_\half$ and $T_\half$ at which the resistivity has
  decreased to half its unitary value [cf.~\Eq{eq:ThalfBhalf}],
  to the scales $\TkFS$ and $\Tksc$ [cf.~\Eqs{eq:chi_m} and
  (\ref{eq:chi_sc})], respectively.  The shown ratios are
  universal numbers of order unity, but not necessarily very
  close to 1, with a significant dependence on $N$: $B_\half/\TkFS =
  1.22, 1.31, 1.60$ and $T_\half/\Tksc = 0.82, 1.02, 1.36$ for $N=1$,
  2 and 3, respectively.  The legend in the lower left of panel (b)
  specifies the NRG parameters used for both panels.}
\label{fig:scaling}
\end{figure}

In numerical work, however, the scaling limit is never fully
realized, since the bandwidth is always finite.  It may thus happen
that a scaling collapse expected analytically is not found when the
corresponding curves are calculated numerically.  For example, if
the Kondo temperature is defined, as seems natural, in terms of a
purely local susceptibility, $\chiloc$, involving only the response of
the local spin to a local field,
\begin{equation}
\frac{4 S(S+1)}{3\pi \Tkloc}  
\equiv 
\chiloc \equiv \frac{d}{d \Bimp}\langle S_z 
\rangle \vert_{B=0} \; , 
\;  
\label{eq:chi_d}
\end{equation}
then curves expected to show a scaling collapse actually do not
collapse onto each other, as pointed out recently in
Ref.~\onlinecite{Hanl2013a} (see Figs.~2(d)-(f) there). That paper
also showed how to remedy this problem: the static spin susceptibility
used to calculate $\Tk$ has to be defined more carefully, and two
slightly different definitions have to be used, depending on the
context. The first option is needed when studying zero-temperature
(i.e.\ ground state) properties as a function of some external
parameter, such as the field dependence of the resisitivity (needed
for $c_B$).
In this case, a corresponding susceptibility defined in
terms of the response of the system's \textit{total} spin to a local
field should be used:
\begin{equation}
\frac{4 S(S+1)}{3\pi \TkFS}  
\equiv 
\chiFS \equiv \frac{d}{d \Bimp}\langle \Sztot 
\rangle \vert_{B=0} 
\; . 
\label{eq:chi_m}
\end{equation}
The superscript FS stands for ``Friedel sum rule'', to highlight the
fact that using this rule to calculate the linear response of $\langle
\Sztot \rangle$ to a local field directly leads to relation
(\ref{eq:Friedel-magnetization}) between $\chiimp$ and $\Tk$.  
The second option is needed
when studying dynamical or thermal quantities that depend
on the system's many-body excitations for given fixed external
parameters (e.g.\ fixed $B=0$), such as the
temperature-dependence of the resistivity (needed for $c_T$), or the
curvature of the Kondo resonance (needed for $c_\varepsilon$).
In this case, one should use
\begin{eqnarray}
\frac{4 S(S+1)}{3\pi \Tksc}  
& \equiv &
\chisc \equiv 
2\chiFS
- \chiloc  \; . 
\label{eq:chi_sc}
\end{eqnarray}
The superscript sc stands for ``scaling'', to indicate that this
definition of the Kondo temperature ensures\cite{Hanl2013a} a scaling
collapse of dynamical or thermal properties.  Figure~\ref{fig:scaling}
demonstrates that a scaling collapse is indeed found when the field-
or temperature-dependent resistivity, plotted versus $B/\TkFS$ or
$T/\Tksc$, respectively, is calculated for two different values of
$\Jk$ (solid and dashed lines, respectively). Note that this works
equally well for $N=1, 2$ and $3$. (For $N=1$, such scaling collapses
had already been shown in Ref.~\onlinecite{Hanl2013a}.) 

We remark that the three Kondo temperatures defined in \Eqs{eq:chi_d}-(\ref{eq:chi_sc})
differ quite significantly from each other for
the Kondo Hamiltonian of \Eq{eq:Kondomodel}, with differences as large
as 12\%, 31\% and 55\% for $N=1$, 2 and 3, respectively, for the
parameters used in Fig.~\ref{fig:scaling}.  This indicates that
although we have chosen bare paramters for which $\Tk/D$ is smaller
than $ 10^{-4}$, we have still not reached the scaling limit [in
which the definitions Eq.~(\ref{eq:chi_d})-(\ref{eq:chi_sc}) of the Kondo temperature should
all coincide numerically\cite{Hanl2013a}]. We have checked that the differences
between $\Tkloc$, $\TkFS$ and $\Tksc$ decrease when $\nu \Jk$ is
reduced in an attempt to get closer to the scaling limit, but estimate
that truly reaching that limit would require $\nu \Jk < 0.01$ for the
Kondo model, implying $\Tk/D < 10^{-45}$. Thus, reaching the scaling
limit by brute force is numerically unfeasible.  Therefore, using
$\TkFS$ and $\Tksc$ rather than $\Tkloc$ is absolutely essential for
obtaining scaling collapses. It is similarly essential for an accurate
determination of the Fermi-liquid parameters. Correspondingly, for the
results discussed below, we have used $\TkFS$ as definition of the
Kondo temperature when extracting $c_B$, and $\Tksc$ when extracting
$c_T$ and $c_\varepsilon$.

\begin{figure*}[t!]
\includegraphics{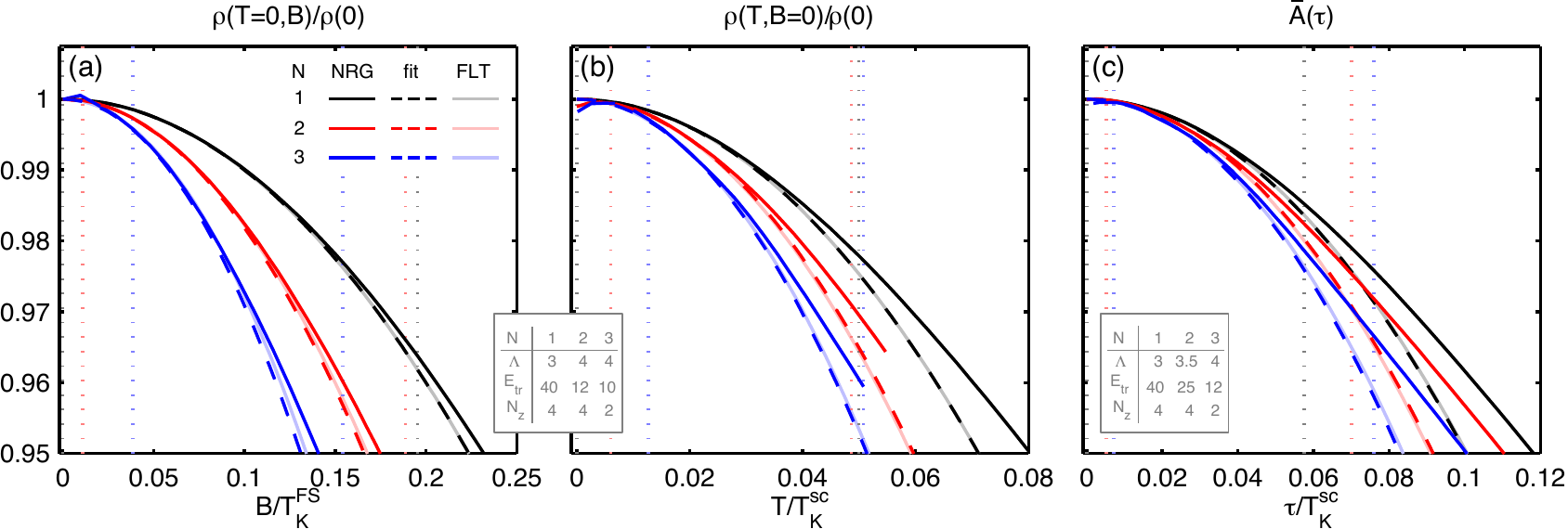}
\caption{(Color online) (a) Resistivity as function of magnetic field
  at $T=0$, (b) resistivity as function of temperature at
    $B=0$, and (c) the weighted spectral function ${\bar A}({\tau})$
  [cf.~\Eq{eq:defI}] at $T=B=0$, all shown for $N=1,2,3$. Each
  panel contains NRG data (heavy solid lines), the quadratic term from
  a fourth order polynomial fit (heavy dashed lines) and the
  corresponding predictions from FLT of \Eq{eq:cBTeresults} for
  the quadratic term (light solid lines). Left and right
    vertical dotted lines in matching colors indicate the lower and
    upper borders of the fitting range used for each $N$.  
    The boxed legends specify the NRG parameters used here.}
\label{fig:FLcoeff}
\end{figure*}

\subsection{Using unbroadened discrete data only}

When one is interested in spectral properties, one typically has to
broaden the discrete data. For the determination of the Fermi-liquid
coefficients, however, where high numerical accuracy is required, it
is desirable to avoid standard broadening. For the calculation of
$c_T$ and $c_B$ this can be achieved\cite{Weichselbaum2007} by
directly inserting the Lehmann sum over $\delta$ functions for the
spectral function $A_{m\sigma}(\varepsilon,T,B)$
[Eq.~(\ref{eq:discdata})] into the energy integral for $\rho(T,B)$
[Eq.~(\ref{eq:G(T,B)})], resulting in a sum over discrete data
points that produces a smooth curve.  The curve is smooth because
\Eq{eq:G(T,B)} in effect thermally broadens the $\delta$ peaks in
the Lehmann representation. This is true even in the limit $T \to
0$, because in NRG calculations it is realized by taking $T$ nonzero, but
much smaller than all other energy scales.

For the determination of $c_{\varepsilon}$, in contrast, one faces the
problem that $A_{m\sigma}(\varepsilon,0,0)$ is represented not as an
integral of a sum over discrete $\delta$ functions, but directly in
terms of the latter. To avoid having
to broaden these by hand, it is desirable to find a way to
extract $c_{\varepsilon}$ from an expression involving an integral
over the discrete spectral data, as for $c_B$ and $c_T$.  This can be
achieved as follows. First, note that $c_{\varepsilon}$ is, by
definition, a coefficient in the general Taylor expansion of the
normalized spectral function $A^{\rm norm}(\varepsilon) \equiv
A_{m\sigma}(\varepsilon ,0,0)/A_{m\sigma}(0,0,0)$ for small
frequencies,
\begin{equation}
A^{\rm norm}(\varepsilon)=\sum_{n=0}^{\infty} a_n (\varepsilon/\Tk)^n, \qquad
c_\varepsilon = a_2 \; . 
\label{eq:AomegaTaylor}
\end{equation}
Due to particle-hole symmetry, $a_n=0$ for all $n$ odd, and by
  definition $a_0=1$.  To determine $a_2$ from an integral over
discrete data, we consider a weighted average of $A^{\rm
  norm}(\varepsilon)$ over $\varepsilon$,
\begin{equation}
{\bar A}(\tau) \equiv \int d\varepsilon A^{\rm norm}(\varepsilon)P_{\tau}(\varepsilon),
\label{eq:defI}
\end{equation}
where $P_{\tau}(\varepsilon)$ is a symmetric weighting function
of width $\tau$ and weight 1, and moments defined by  
\begin{equation}
\int d\varepsilon (\varepsilon/{\tau})^n P_{\tau}(\varepsilon) \equiv p_n  
\label{eq:defmoments}
\end{equation}
for integer $n \ge 0$ (with $p_0 = 1$). Here we use 
\begin{equation}
P_{\tau}(\varepsilon)=\frac{1}{4\tau}\frac{1}{\cosh^2\left( \varepsilon/2\tau \right)}=-\frac{\partial f(\varepsilon, \tau)}{\partial \varepsilon} , 
\label{eq:deffprime}
\end{equation}
but other choices are possible, too (e.g., a Gaussian peak).  Clearly,
the leading ${\tau}$ dependence of ${\bar A}({\tau})$ for small
${\tau}$ reflects the leading ${\varepsilon}$ dependence of $A^{\rm
  norm}(\varepsilon)$ and allows for an accurate determination of
$a_2$.  Indeed, using \Eqs{eq:AomegaTaylor}-(\ref{eq:deffprime}),
we obtain a power-series expansion for ${\bar A}({\tau})$ of the form
${\bar A}({\tau})=\sum_{n} a_n p_n {(\tau/\Tk)}^n$.
Thus, by fitting 
${\bar A}^{\rm fit}({\tau})=\sum_n f_n {\tau}^n$ 
to the NRG data for ${\bar A}({\tau})$, one can determine the desired
coefficients in (\ref{eq:AomegaTaylor}) using $a_n=\Tk^n f_n/p_n$. In
particular, the Fermi-liquid coefficient of present interest is given
by $c_{\varepsilon}=a_2=\Tk^2 f_2/p_2$.

\subsection{Extraction of Fermi-liquid coefficients}

Figs.~\ref{fig:FLcoeff}(a)-(c) show our NRG data (heavy solid lines) for the
resistivity plotted versus $B/\TkFS$ at zero temperature or plotted
versus $T/\Tksc$ at zero field, and for the weighted spectral function
plotted versus $\tau /\Tksc$, respectively.  We determined the
Fermi-liquid coefficients $c_B$, $c_T$ and $c_{\varepsilon}$ from the
quadratic terms of fourth-order polynomial fits to these
curves. Including the fourth-order term allows the fitting range to be
extended towards somewhat larger values of the argument, thus
increasing the accuracy of the fit.  For each solid curve, the
quadratic term from the fit is shown by heavy dashed lines; these are
found to agree well with the corresponding predictions from FLT,
shown by light lines of matching colors. The level of agreement
is quite remarkable, given the rather limited range in which the
behavior is purely quadratic: with increasing argument, quartic
contributions become increasingly important, as reflected by the
growing deviations between dashed and solid lines; and at very small
values of the argument ($\lesssim 0.02$), the NRG data become
unreliable due to known NRG artefacts.

Numerical values for the extracted Fermi-liquid coefficients are given
in Table \ref{tab:NRG_results}; they agree with those predicted
analytically to within $\leq 5\%$. This can be considered excellent
agreement, especially for the numerically 
very challenging case of $N=3$. 

\begin{table}
\vspace{0.5cm}
\begin{center}
\setlength\extrarowheight{7pt}
\setlength{\tabcolsep}{10pt}
\begin{tabular}{c c c c}
\hline\hline
$N$ 
& $c_B^{\rm NRG}/c_B^{\FLT}$  & $c_T^{\rm NRG}/c_T^{\FLT}$ & 
$c_{\varepsilon}^{\rm NRG}/c_{\varepsilon}^{\FLT}$ 
\\\hline
1   &                    $1.00 \pm 0.01$ &                    $1.00 \pm 0.01$ & $1.01 \pm 0.03$ \\
2   &                    $1.02 \pm 0.03$ &                    $0.98 \pm 0.03$ & $0.99 \pm 0.03$ \\
3   &                    $1.05 \pm 0.05$ &                    $1.01 \pm 0.03$ & $1.02 \pm 0.07$ \\
\hline\hline
\end{tabular}
\caption{Numerically extracted values of $c_B$, $c_T$ and
  $c_{\varepsilon}$, given here relative to the corresponding
  predictions from FLT of \Eq{eq:cBTeresults}.  The deviations
  between NRG and FLT values are $\leq 5\%$ in all cases.  To
  numerically determine these coefficients, we used the quadratic
  coefficient of a fourth-order polynomial fit to the corresponding NRG
  data. Error bars were estimated by comparing the quartic fits
  to polynomial fits of different higher orders.}
\label{tab:NRG_results}
\end{center}
\end{table}
 
\section{Conclusions}
\label{sec:conclusions}

Our two main results can be summarized as follows. First, we have
presented a compact derivation of three Fermi-liquid coefficients for
the fully-screened $N$-channel Kondo model, by generalizing
well-established calculations for $N=1$ to general $N$.  The
corresponding calculations, building on ideas of Nozi\`eres, Affleck
and Ludwig, and Pustilnik and Glazman, are elementary. We hope that
our way of presenting them emphasizes this fact, and perhaps paves the
way for similar calculations in less trivial quantum impurity problems
that also show Fermi-liquid behavior, such as the asymmetric
single-impurity Anderson Hamiltonian, or the 0.7-anomaly in quantum
point contacts.\cite{Bauer2013}

Second, we have established a benchmark for the quality of NRG results
for the fully screened $N$-channel Kondo model, by showing that it is
possible to numerically calculate equilibrium Fermi-liquid
coefficients with an accuracy of better than 5\% for $N=1$, 2 and
3. To achieve numerical results of this quality, two technical
ingredients were essential, both of which became available only
recently: first, exploiting larger-rank non-Abelian symmetries
in the numerics;\cite{Weichselbaum2012b,Moca2012} and second,
carefully defining the Kondo temperature\cite{Hanl2013a} in such a way
that numerically-calculated universal scaling curves are indeed
universal, in the sense of showing a proper scaling collapse, despite
the fact that the scaling limit $\Tk/D \to 0$ is typically not
achieved in numerical work.

\section*{Acknowledgements}
We acknowledge helpful discussions with K.~Kikoin, C.~Mora,
A.~Ludwig and G.~Zar\`and. We are grateful to D.~Schuricht
for drawing our attention to Ref.~\onlinecite{H10}, and for
sending us a preprint of Ref.~\onlinecite{Horig2014} prior to its
submission. The latter work, which we received in the final stages of
this work, also uses $H_\lambda$ of \Eq{eq:AL-start-JJ} as starting
point for calculating Fermi-liquid coefficients for the $N$-channel
Kondo model, and its result for $c_{T}$ is consistent with our own.
We gratefully acknowledge financial support from the DFG (WE4819/1-1
for A.W., and SFB-TR12, SFB-631 and the Cluster of Excellence
Nanosystems Initiative Munich vor J.v.D., M.H., and A.W.)

\appendix*
\section{} 

\label{sec:AppendixA}
This appendix offers a pedagogical derivation of the Hamiltonian
$H_\lambda$ given in \Eq{eq:H-momentum-main} of the main text using the
point-splitting regularization strategy, following AL (Appendix~D of
\onlinecite{AL93}). Its main purpose is to show how the relation
$\alpha = 3 \psi \nu = 1/\Tk$ between Fermi-liquid parameters that NB had found
by intuitive arguments\cite{Nozieres1980} follows simply and naturally
from point splitting. For a detailed discussion of the point-splitting
strategy, see Refs.~\onlinecite{Affleck1986,L92,vonDelft1995}.

According to AL, the leading
irrelevant operator for the fully screened $N$-channel Kondo
model has the form 
\begin{eqnarray}
H_{\lambda}=-\lambda : \! \vec J(0)\cdot \vec J(0) \! : .
\label{eq:AL-startingpoint}
\end{eqnarray}
Here $\vec J(x) =\sum_{m=1}^{N} : \! \vec J_m  (x) \! :$
is the total (point-split) spin density from all channels at 
position $x$ (the impurity or dot sits at $x= 0$), 
and 
\begin{equation}
  \vec J_m(x)=\frac{1}{2}\sum_{\sigma\sigma'} 
\Psi^{\dagger}_{m\sigma}(x){\vec \tau}_{\sigma\sigma'} \Psi_{m\sigma'}(x)
\label{eq3}
\end{equation}
is the corresponding (non-point-split) spin density for
channel $m$.  Here $:...:$ denotes point splitting,
\begin{eqnarray}
  \label{eq:point-splitting}
  :\!  A(x) B(x)\! : \equiv \lim_{\eta \to 0} 
\Bigl[ A(x+\eta) B(x) - \bcontraction[0.6ex]{}{A}{(x+\eta)}{B}A(x+\eta)B(x) 
\Bigr] , \qqph
\end{eqnarray}
a field-theoretic scheme for 
regularizing products of operators
 at the same point by subtracting their
ground state expecation value, $\bcontraction[0.6ex]{}{A}{}{B}AB = 
 \langle  A B\rangle$. (In most cases, point splitting
is equivalent to normal ordering.) 
For present purposes, we follow AL\cite{AL93} and take 
\begin{eqnarray}
  \label{eq:defineFermions}
\Psi_{m\sigma}(x) = \frac{1}{\sqrt L} \sum_{k} e^{- i k x} \psi_{km\sigma}   
\end{eqnarray}
to be free fermion fields 
with linear dispersion ($\xi_k = k \hbar \vF $) in a box of length
$L\to \infty$ (with $k\in 2 \pi n / L$, $n \in \mathbb{Z}$),
with normalization $\{\psi_{km\sigma}, \psi^\dagger_{k'm'\sigma'}\} = 
\delta_{kk'}\delta_{mm'}\delta_{\sigma \sigma'}$ and 
free ground state correlators
\begin{equation}
\langle \Psi^\dagger_{m\sigma}(x)
\Psi_{m'\sigma'}(0)\rangle=
\langle \Psi_{m\sigma}(x)
\Psi^\dagger_{m'\sigma'}(0)\rangle=
\frac{\delta_{mm'}\delta_{\sigma\sigma'}}{2 \pi i x} \; .
\end{equation}
Note that we follow PG in our choice of field normalization, which differs
from that used by AL\cite{AL93} by $\Psi_{\rm here} = \psi_{\rm AL} /
\sqrt {2 \pi}$.  Consequently, our coupling constant is related to
theirs by $\lambda_{\rm here} =(2 \pi)^2 \lambda_{\rm AL}$.

In the definition of $H_\lambda$, point splitting is needed because the
product of two spin densities, $\vec J(x+\eta)\cdot \vec J(x)$,
diverges with decreasing seperation $\eta$ between their arguments.
To make this explicit, we use Wick's theorem,
\begin{eqnarray}
: \! \! AB \! :  : \! CD \! : \, = \, 
: \!\!  AB CD \! : \! + \! 
: \! \!  \bcontraction[0.6ex]{A}{B}{}{C} AB CD \! : \! + \! 
: \! \! \bcontraction[0.6ex]{}{A}{BC}{D} AB CD \! : \! + \! 
: \! \!   \bcontraction[0.6ex]{A}{B}{}{C} \bcontraction[1ex]{}{A}{BC}{D} AB CD \! : ,
\nonumber 
\end{eqnarray}
 to rewrite the product of spin densities as follows: 
\begin{widetext}
  \begin{subequations}
\begin{eqnarray}
\vec J(x+\eta)\cdot \vec J(x) & = &  
\frac{1}{4}\sum_{m\sigma \sigma'}\sum_{m' \bsigma\bsigma'}
:\! 
\Psi^{\dagger}_{m\sigma}(x+\eta){\vec \tau}_{\sigma\sigma'}\Psi_{m\sigma'}(x+\eta) 
\! : \, : \! 
\Psi^{\dagger}_{m'\bsigma}(x){\vec \tau}_{\bsigma\bsigma'}\Psi_{m'\bsigma'}(x) 
\! : 
\label{eq:Wick1} 
\\ \nonumber
& = &  \frac{1}{4}\sum_{m\sigma \sigma'}\sum_{m' \bsigma\bsigma'}
{\vec \tau}_{\sigma\sigma'}\cdot{\vec \tau}_{\bsigma\bsigma'}
\Biggl[:\! \Psi^\dagger_{m\sigma}(x+\eta)\Psi_{m\sigma'}(x+\eta)
\Psi^\dagger_{m'\bsigma}(x)\Psi_{m'\bsigma'}(x) \! : \Biggr. 
\\
& & \Biggl. 
+\frac{\delta_{mm'}}{2 \pi i \eta}\Bigl(
\delta_{\sigma'\bsigma}:\! \Psi^\dagger_{m\sigma}(x+\eta)\Psi_{m\bsigma'}(x)\! :
{} + {} \delta_{\sigma\bsigma'}:\! \Psi_{m\sigma'}(x+\eta)\Psi^\dagger_{m\bsigma}(x)\! :\Bigr)
 + \frac{\delta_{\sigma\bsigma'}\delta_{\sigma'\bsigma}
\delta_{mm'}}{(2 \pi i \eta)^2} \Biggr] \; . \qqph 
\label{eq:Wick}
\end{eqnarray}
\end{subequations}
The point-splitting prescription in \Eq{eq:AL-startingpoint} subtracts
off the $1/\eta^2$ divergence of the last term of \Eq{eq:Wick}. The
contributions of the second and first terms to $H_\lambda$ can be
organized as $H_{\lambda} = \Hone + \Hin$, describing
single-particle elastic scattering and two-particle interactions,
respectively. Taking $x=0$ and $\eta \to 0$, we find:
\begin{subequations}
\begin{eqnarray}
  \label{eq:H1xixi}
  \Hone & = &  
- \frac{\lambda}{8 \pi i } \lim_{\eta \to 0} \sum_{m \sigma\sigma'} 
: \! \frac{1}{\eta} \Bigl[ \Psi^\dagger_{m\sigma}(\eta) \vec \tau^2_{\sigma \sigma'}\Psi_{m\sigma'}(0)
-  \! \Psi^\dagger_{m\sigma'}(0) \vec \tau^2_{\sigma' \sigma}
\Psi_{m\sigma}(\eta) \Bigr] \! :
\label{eq:Hel-1} \\
& = & - \frac{3 \lambda}{8 \pi i}  \lim_{\eta \to 0} \sum_{m \sigma }
: \! \left[ \frac{1}{\eta} \Bigl(\Psi^\dagger_{m\sigma}(\eta) - 
\Psi^\dagger_{m\sigma}(0) \Bigr) \Psi_{m\sigma}(0) 
-  \Psi^\dagger_{m \sigma}(0)\frac{1}{\eta} \Bigl(  \Psi_{m\sigma}(\eta) - 
\Psi_{m\sigma}(0) \Bigr)  \right]  \! : \; 
\label{eq:Hel-2} \\
& = &  
- \frac{ 3\lambda}{8 \pi i } \sum_{m\sigma}
 : \! \Bigl[
\bigl( \partial_x \Psi^{\dagger}_{m\sigma} \bigr) (0) \Psi_{m\sigma} (0) 
- \Psi^{\dagger}_{m\sigma} (0) \bigl(\partial_x \Psi_{m\sigma})(0) \Bigr] \! : 
, 
\label{eq:Hel-3}
\end{eqnarray}
\end{subequations}
\end{widetext}
\begin{eqnarray}
\Hin  =-\lambda \sum_{m m'}: \! \vec J_m (0)\cdot \vec J_{m'} (0) \! : .
\label{eq7}
\end{eqnarray}
To obtain \Eq{eq:Hel-2}, we used $\vec\tau^2_{\sigma\sigma'}=3\delta_{\sigma\sigma'}$
and subtracted and added :$\Psi^\dagger_{m\sigma}(0) \Psi_{m\sigma}(0)$:
inside the square brackets. 
Now pass to the momentum representation, using \Eq{eq:defineFermions}
and the shorthand notations (following PG\cite{PG})
\begin{subequations}
\begin{align}
  \rho_{m\sigma} (0) & = \frac{1}{L} \rho_{m\sigma}, & \rho_{m\sigma}
  & = \sum_{kk'} \psi^\dagger_{km\sigma} \psi_{k'm\sigma} \; ,
  \\
  \vec J_m (0) & = \frac{1}{L} \vec j_m , & \vec j_m & = \frac{1}{2}
  \sum_{kk'\sigma\sigma'} \psi^\dagger_{km\sigma} \vec
  \tau_{\sigma\sigma'} \psi_{k'm\sigma'} \; ,
 \end{align} 
\end{subequations}
for the conduction electron 
channel-$m$ charge and spin densities at
the impurity. This gives 
\begin{subequations}
\begin{eqnarray}
  \label{eq:Hel-momentum}
 \Hone & = & - \frac{\alpha_1}{2 \pi \nu}
\sum_{m\sigma kk'} (\xi_k + \xi_{k'}) 
: \! \psi^\dagger_{km\sigma}   \psi_{k'm\sigma} \! : \; , \quad
\phantom{.}
\\
\Hin & = & - \frac{2\phi_1}{3 \pi \nu^2}
\sum_{mm'} : \! \vec j_m \cdot \vec j_{m'} \! : \;  .
\label{eq:Hint-momentum}
\end{eqnarray}
\end{subequations}
Here $\nu = L/(2 \pi \hbar \vF)$ is
the extensive 1D density of states per spin and channel, 
and the prefactors were expressed in terms of the constants
\begin{eqnarray}
\label{eq:identify-alpha1-phi1}
\alpha_1 = \phi_1 = \frac{3 \lambda}{8 \pi(\hbar \vF)^2} = \frac{1}{\Tk} 
\; . 
\end{eqnarray}
(This notation is consistent 
with that of Ref.~\onlinecite{Horig2014}, where
$H_\lambda$ served starting point for calculating
Fermi-liquid corrections, too.)
Checking dimensions, with $[H_\lambda]$=$\mathcal{E}$ and 
$[\Psi_{m\sigma}]$=$1/\sqrt{\mathcal{L}}$
($\mathcal{E}$ stands for energy, $\mathcal{L}$ for length),
we see that $[\lambda]$=$\mathcal{EL}^2$. Since 
$[\nu]$=$1/\mathcal{E}$, $[\hbar \vF]$=$\mathcal{EL}$, 
we have $[\alpha_1]= [\phi_1] = 1/\mathcal{E}$,
thus, $\alpha_1$ and $\phi_1$ have dimensions of inverse energy. In
the main text, they are identified with $1/\Tk$; in fact, the
numerical prefactor in \Eq{eq:identify-alpha1-phi1} is purposefully
chosen such that the leading term in the expansion
(\ref{eq:GP-phase-shift-a-la-NB}) of the phase shift
$\tdelta_{m\sigma}(\varepsilon)$ turns out to take the form $\varepsilon/\Tk$.

To elucidate how the case $N>1$ differs from $N = 1$,
we write $\Hin = \Htwo + \Hthree$ in the main text, with 
$\Htwo$ and $\Hthree$ given in \Eqs{eq:Htwo-momentum}
and  (\ref{eq:Hthree-momentum}), 
respectively, where $\Hthree$ occurs only for $N>1$.

\end{document}